\documentclass[%
 reprint,
superscriptaddress,
 amsmath,amssymb,
 aps,
pra,
]{revtex4-2}

\usepackage{graphicx}
\usepackage{braket}
\usepackage{xcolor}
\usepackage{comment}
\usepackage{booktabs}
\usepackage[normalem]{ulem}

\begin{document}

\title{Real-time probing of quadrupolar contributions to the absorption cross section of non-periodic systems}

\author{Anvar Khujakulov}
\email{anvar.khujakulov@uni-jena.de}
\affiliation{Friedrich-Schiller Universit\"at Jena, Institute for Condensed Matter Theory and Optics, 07743 Jena, Germany}

\author{Michele Guerrini}
\affiliation{Friedrich-Schiller Universit\"at Jena, Institute for Condensed Matter Theory and Optics, 07743 Jena, Germany}

\author{Carlo Andrea Rozzi}
\affiliation{CNR -- Istituto Nanoscienze, 41125 Modena, Italy}

\author{Caterina Cocchi}
\email{caterina.cocchi@uni-jena.de}
\affiliation{Friedrich-Schiller Universit\"at Jena, Institute for Condensed Matter Theory and Optics, 07743 Jena, Germany}
\affiliation{Abbe Center of Photonics, Friedrich-Schiller-Universit\"at Jena, 07745, Jena, Germany}

\date{\today}

\begin{abstract}
The non-perturbative evaluation of multipolar cross sections is essential for probing atomic and molecular responses to spatially inhomogeneous electric fields characteristic of nanoscale environments where the conventional dipole approximation breaks down. Taking the hydrogen atom as an analytical and numerical benchmark, we map in real time quadrupole interactions driven by instantaneous electric field gradients. Our study explores two distinct interaction regimes.  Under a uniform field, quadrupolar responses are activated by multi-step dipole transitions, responsible for sub-1~eV excited-state absorption. Conversely, under pure spatial gradients, the weak-field response is dominated by the single-photon $1s \to 3d$ quadrupole resonance at 12.1~eV, while strong gradients induce low-energy stimulated emission via transiently driven coherent populations. This dynamical analysis is complemented by a rigorous evaluation of the symmetry aspects of the quadrupole response and the applied electric field. This methodology establishes the theoretical and computational foundation for future non-perturbative multipole simulations of molecules and nanostructures in the near-field regime.
\end{abstract}

\maketitle


\section{Introduction}\label{sec:intro}
Recent breakthroughs in nanophotonics, tip-enhanced near-field microscopies, and plasmonic nanocavities has fundamentally altered the spatial scales of modern light-matter spectroscopy~\cite{verm17cr,li+17csr,shi+17cr,rich+17csr,macc+21na}. These established techniques enable the confinement of intense, ultrafast electromagnetic fields into sub-nanometer hot spots~\cite{rope+07njp,bhat+17cc}. In highly structured environments, the spatial distribution of the electric field varies dramatically~\cite{raza+15jpcm,zhu+16natcom}, leading to a complete breakdown of the dipole approximation~\cite{yama+15pra}, while simultaneously activating higher-order multipolar interactions that shape the response of the system under these conditions~\cite{iwas+16jcp}.

The electric quadrupole interaction constitutes the leading correction to the multipolar expansion~\cite{ewig+02jpca,holt-karl08jcc} and imposes angular-momentum selection rules ($\Delta \ell =0,2$) activating strictly forbidden transitions in conventional far-field spectroscopies. Although quadrupole effects are formally well understood in static atomic and molecular environments~\cite{Varshalovich1988,johnson2007}, and pioneering implementations of these terms have been recently proposed for time-dependent density functional theory (TDDFT)~\cite{vang+99cpc}, their role in ultrafast dynamics remains comparatively unexplored. Historically, quadrupole contributions have been treated as negligible corrections or implicitly folded into nonlinear dipole responses~\cite{kris-prak66rmp}. However, under intense illumination or sharp spatial gradients, these channels can dominate the time-evolution of the quantum system, acting as an independent control knob on equal footing with field intensity and carrier frequency~\cite{yann-pasp15jmo,iwas24jpcl,loul+25acsphot}.

Real-time electronic-structure methods provide a natural and computationally elegant platform for modeling non-uniform perturbations~\cite{andr+07jcp,taki+07jcp,gonc-varg12jcp}. Within this framework, the impulsive perturbation method (known as $\delta$-kick) introduced by Yabana and Bertsch~\cite{quad:Yabana1996} enables the broadband mapping of linear~\cite{thie-kuem09pccp,tuss+15jctc} and non-linear absorption cross sections~\cite{cocchi2014ab,guandalini2021,quad:Driouech2025} within a single time-propagation, avoiding cumbersome calculation of high-order response functions. While fully self-consistent minimal-coupling frameworks paired with Maxwell solvers~\cite{list2020,bonafe2025full,bonafe2025} offer a general route to beyond-dipole physics, they introduce substantial theoretical and computational complexity. Alternatively, real-time TDDFT (RT-TDDFT)~\cite{rung-gross84prl} implemented on real-space grids~\cite{andrade2015} provides a highly efficient alternative, free from basis-set incompleteness errors that plague localized atomic orbital representations~\cite{beck00rmp,kuem-kron08rmp}.

In this work, we develop a non-perturbative multipole formalism within the RT-TDDFT code \texttt{Octopus}~\cite{tancogne2020} and validate its accuracy against analytical and numerical solutions of the time-dependent Schr\"odinger equation (TDSE). Taking the hydrogen atom as a spherically symmetric, analytically tractable benchmark~\cite{kim-kim18nanophotonics}, we isolate the fundamental light-matter interaction from the structural anisotropies of molecular geometries and systematically explore four representative scenarios: conventional weak-field dipole transitions as a reference, far-field excitations driven by strong intensities, and quadrupole interactions triggered by both weak and intense spatial field gradients. 

Our real-time simulations reveal distinct dynamical behaviors across these regimes. In the uniform strong-field domain, quadrupolar transitions are effectively triggered by multi-step dipole absorption sequences, leading to prominent sub-1~eV excited-state absorption features. Conversely, under spatial field gradients where the field vanishes at the nucleus, the linear response is dominated by the single-photon $1s \to 3d$ quadrupole resonance at 12.1~eV. In the strong-gradient regime, the spectrum yields positive and negative peaks below 1~eV due to stimulated emission from transiently pumped, coherently excited populations. Finally, we resolve geometric aspects of the Cartesian quadrupole response under spatial rotation, showing that, while individual diagonal components exhibit artificial amplitude splitting due to the traceless constraint, absolute rotational covariance is perfectly preserved within the coordinate-swapped off-diagonal subspace. Validated by the analytical benchmark of the hydrogen atom, this work provides the physical and computational foundation for the future extension of non-perturbative multipole RT-TDDFT simulations to molecules and nanostructures in the near-field excitation regime.

\section{Theoretical Framework}\label{sec:theory}
To model quadrupole interactions in spatially non-uniform fields, we consider a hydrogenic system interacting with an external electric perturbation. The electron dynamics are governed by the TDSE:
\begin{equation}
  i\hbar\frac{\partial \psi(\mathbf{r},t)}{\partial t}= \left[\hat{H}_0(\mathbf{r})+\hat{H}_{\rm int}(\mathbf{r},t)\right]\psi(\mathbf{r},t),  
  \label{eq:tdse}
\end{equation}
where
\begin{equation}
\hat{H}_0 = -\dfrac{\hbar^2}{2m_e}\nabla^2 - \dfrac{e^2}{4\pi\varepsilon_0 |\mathbf{r} - \mathbf{r}_0|}
\label{eq:field-free_H}
\end{equation}
is the field-free Hamiltonian for a (fixed) nucleus located at position $\mathbf{r}_0$.
For a system exposed to a spatially dependent electric field $\mathbf{E}(\mathbf{r},t)$, a Taylor expansion of the interaction potential around $\mathbf{r} = \mathbf{r}_0$ yields the interaction Hamiltonian $\hat{H}_{\rm int}$. Since the Cartesian quadrupole operator $\hat{Q}_{ij}$ is inherently symmetric under index permutation, it couples exclusively to the symmetric part of the electric field gradient tensor, $E_{ij}(\mathbf{r}_0,t) = \frac{1}{2} \left.\left( \partial_i E_j + \partial_j E_i \right)\right|_{\mathbf{r}=\mathbf{r}_0}$. Retaining terms up to the quadrupole order yields:
\begin{equation}
\hat{H}_{\rm int}(\mathbf{r},t) = \underbrace{-\mathbf{d} \cdot \mathbf{E}(\mathbf{r}_0,t)}_{\hat{H}_{\rm d}} - \underbrace{\frac{1}{3} \sum_{i,j} \hat{Q}_{ij} E_{ij}(\mathbf{r}_0,t)}_{\hat{H}_{\rm qd}},
\label{eq.Hint}
\end{equation}
where the factor 1/3 in the quadrupole term reflects the contraction over this symmetric tensor, which accounts for a factor of 2 relative to the original 1/6 coefficient in the Taylor expansion.
The dipole operator is defined as $\mathbf{d} = e(\mathbf{r}-\mathbf{r}_0)$, and the quadrupole interaction is mediated by the traceless second-rank Cartesian tensor:
\begin{equation}\label{eq.quadrO}
\hat{Q}_{ij} = e\left[(x_i - x_{0,i})(x_j - x_{0,j}) - \frac{1}{3}\delta_{ij}|\mathbf{r} - \mathbf{r}_0|^2\right].
\end{equation}
In contrast to the standard dipole approximation, which assumes a spatially homogeneous field, the inclusion of $\hat{H}_{\rm qd}$ captures the effects of local field gradients ($\partial_j E_i$), which become significant under spatial field confinement and at high intensities. 

We model the excitation via an impulsive, spatially non-uniform electric field:
\begin{equation}
    \mathbf{E}(\mathbf{r},t) = \mathbf{E}_0(\mathbf{r})\delta(t),
\end{equation}
corresponding to the Yabana-Bertsch $\delta$-kick~\cite{quad:Yabana1996}. The time-dependent wavefunction is expanded onto the eigenstates of the field-free Hamiltonian [Eq.~\eqref{eq:field-free_H}], with corresponding eigenvalues $\varepsilon_j$:
\begin{equation}
    \psi(\mathbf{r},t) = \sum_{j} c_j(t) e^{-i\varepsilon_j t / \hbar} \phi_j(\mathbf{r}).
    \label{eq.psif}
\end{equation}
For a system initially in the ground state $\phi_0$, the transition coefficients immediately following the pulse ($t > 0$) are evaluated from perturbation theory as:
\begin{equation}\label{eq.c_coef_refined}
c_j = \langle \phi_j | e^{-i \hat{H}_{\rm int}/\hbar} | \phi_0 \rangle \approx \delta_{j,0} - \frac{i}{\hbar} \langle \phi_j | (\hat{H}_{\rm d} + \hat{H}_{\rm qd}) | \phi_0 \rangle + \dots 
\end{equation}
The explicit forms of these scalar coefficients across different interaction regimes are listed in Table~\ref{tab:regimes}.

\begin{table*}
\centering
\renewcommand{\arraystretch}{1.4} 
\setlength{\tabcolsep}{10pt}
\begin{tabular}{lllll}
\hline\hline
\textbf{Label} & \textbf{Field Strength} & \textbf{Interaction} & \textbf{Condition} & \textbf{Coefficient ($j > 0$)} \\
\hline
 WD & Weak   & Dipole     & $H_{\rm d} \ll 1$                  & $c_j \approx -i H_{\rm d}$ \\
 SD & Strong & Dipole     & $H_{\rm d} \gtrsim 1$               & $c_j = -i H_{\rm d} - \frac{1}{2}H_{\rm d}^2 + \dots$ \\
 WD+WQ & Weak   & Dipole+Quadrupole & $H_{\rm d} + H_{\rm qd} \ll 1$          & $c_j \approx -i (H_{\rm d} + H_{\rm qd})$ \\
 SD+SQ & Strong & Dipole+Quadrupole & $H_{\rm d} + H_{\rm qd} \gtrsim 1$       & $c_j = -i (H_{\rm d} + H_{\rm qd}) - \frac{1}{2}(H_{\rm d} + H_{\rm qd})^2 + \dots$ \\
\hline\hline
\end{tabular}
\caption{Excitation regimes and transition coefficients $c_j(t>0)$ in Eq.\eqref{eq.c_coef_refined} computed under an impulsive $\delta$-kick excitation in atomic units, distinguishing between weak/strong field strengths and dipole/quadrupole interactions. The terms $H_{\rm d}$ and $H_{\rm qd}$ denote the dipole $\langle \phi_j | \hat{H}_{\rm d} | \phi_0 \rangle$ and quadrupole $\langle \phi_j | \hat{H}_{\rm qd} | \phi_0 \rangle$ transition matrix elements, respectively.}
\label{tab:regimes}
\end{table*}

The threshold field strength $E_{\text{th,d}}$ separating the weak dipole (WD) and strong dipole (SD) regimes can be evaluated from the transition matrix elements between the ground state and the first accessible excited state. For the hydrogen atom, this corresponds to the $1s \to 2p$ transition along the polarization axis $\hat{z}$:
\begin{equation}\label{eq.dipH}
\lvert \braket{1s | \hat{d}_z | 2p} \rvert= 0.745 \, e a_0 \sim e a_0,    
\end{equation}
where $\hat{d}_z = -e \hat{z}$ is the dipole operator along the polarization axis and $a_0$ is the Bohr radius, leading to the relation:
\begin{equation}\label{eq.dipH_ham}
|H_d| \sim e a_0 E_0.
\end{equation}
In atomic units (a.u., $\hbar = e = m_e = 1$), the condition $H_{\rm d} \sim 1$ from Table~\ref{tab:regimes} simplifies to:
\begin{equation}\label{eq.dipH_F}
E_{\text{th,d}} \sim 1 \ \text{a.u.}
\end{equation}
corresponding to ${E_{\text{th,d}} \sim 5 \times 10^{11} \ \text{V/m}}$ in SI units.

To evaluate the quadrupole contributions, we parameterize the spatial profile of the field amplitude as:
\begin{equation}\label{eq.Elfi}
E_i(\mathbf{r}) = E_{0} \left( \alpha_i + \beta_i x + \gamma_i y + \delta_i z \right) \;\; \text{with} \;\; i \in \{x, y, z\},
\end{equation}
where $\alpha_i$ is dimensionless, while the gradient coefficients $\{\beta_i, \gamma_i, \delta_i\}$ carry the dimension of an inverse length. To satisfy Maxwell's equations in vacuum, the field gradient tensor must remain traceless, enforcing the divergence-free condition $\beta_x + \gamma_y + \delta_z = 0$. We emphasize that, in all real-time simulations, the full spatially dependent field of Eq.~\eqref{eq.Elfi} is coupled to the electron and propagated non-perturbatively without truncation. The multipole expansion of Eq.~\eqref{eq.Hint} is used only to classify and interpret the resulting dipole and quadrupole response channels, rather than to approximate the propagated Hamiltonian.

The spatial field gradient introduces quadrupole-allowed transitions mediated by $\hat{H}_{\rm qd}$, which are governed by the selection rules $\Delta \ell =0,\pm 2$ (excluding the  forbidden $0 \rightarrow 0$ transition) and $\Delta m = 0, \pm 1, \pm 2$~\cite{quad:fam18}. The corresponding components of the time-dependent Cartesian quadrupole tensor are computed as:
\begin{equation}\label{eq.OmeQ}
\Omega^{\rm qd}_{ij}(t) = \langle \psi(t) | \hat{Q}_{ij} | \psi(t) \rangle.
\end{equation}
While $\Omega^{\rm qd}_{ij}(t)$ can develop a non-zero transient expectation value even under spatially uniform strong fields, such configurations do not lead to linear energy absorption unless actively coupled to an external field gradient. 

To evaluate the corresponding thresholds in spatially inhomogeneous fields, we consider the joint dipole and quadrupole interaction terms from Eq.~\eqref{eq.Hint}. The characteristic scale of the quadrupole coupling is governed by the bound-to-bound transition matrix element, an intrinsic atomic quantity of the order of $e a_0^2$. For the hydrogen atom, the relevant matrix element is $|\braket{1s|\hat{Q}_{zz}|3d}| = 1.55\, e a_0^2 \sim e a_0^2$ (Table~\ref{tab:qdt_com}), consistent with the ground-state expectation value $\braket{r^2} = 3\, a_0^2$. Parameterizing the characteristic strength of the electric field gradient as $\partial_j E_i \sim E_0\,\Gamma_0$ from Eq.~\eqref{eq.Elfi} (here, $\Gamma_0$ denotes the electric-field gradient strength, see Appendix.~\ref{app:linear_avg}), the quadrupole interaction energy scales as $|H_{qd}| \sim e\braket{r^2} E_0\,\Gamma_0$. Consequently, the strong-field threshold criterion $H_{\rm d} + H_{\rm qd} \sim 1$ from Table~\ref{tab:regimes} takes the form:
\begin{equation}\label{eq.Hdip_qd}
  e a_0 E_0 + e\braket{r^2} E_0\,\Gamma_0 \sim 1,
\end{equation}
where the first term is the order-of-magnitude dipole interaction of Eq.~\eqref{eq.dipH_ham}.
Expressed in atomic units, where $\braket{r^2} = 3$, this relation yields a generalized threshold field strength:
\begin{equation}\label{eq.thr_FQ}
  E_{\rm th,d+qd}\sim \frac{1}{a_0+\braket{r^2}\,\Gamma_0} = \frac{1}{1+3\,\Gamma_0}.
  \end{equation}
  
In the limit of a spatially homogeneous field ($\Gamma_0 \rightarrow 0$), Eq.~\eqref{eq.thr_FQ} reduces to the pure dipole threshold of Eq.~\eqref{eq.dipH_F}. Conversely, when the uniform dipole component vanishes, ($\alpha_i = 0$ in Eq.~\eqref{eq.Elfi}), the threshold field strength scales inversely with the field gradient:
\begin{equation}\label{eq.thr_FQ1}
  E_{\rm th,qd}\sim \frac{1}{\braket{r^2}\,\Gamma_0} \sim \frac{1}{3\,\Gamma_0}.
  \end{equation}
In this case, the interaction regimes labeled as "WD+WQ" ("SD+SQ") are reduced to just WQ (SQ), as shown in Table \ref{tab:regimes}.
This threshold estimate, paired with the cross-section analysis in Eq.~\eqref{eq:alpha_single_calc_app}, establishes a quantitative boundary for the interaction regimes. In the linear domain, the quadrupole cross section remains independent of, or only weakly sensitive to, variations in $\Gamma_0$.

The primary physical observable extracted from real-time simulations is the photo-absorption cross-section $\sigma_{\rm tot}(\omega)$. To model isotropic gas-phase conditions without the prohibitive computational cost of explicitly averaging over random molecular orientations, we project the tensorial response onto the rotationally invariant ($\ell=0$) component of its $SO(3)$ irreducible representation. For spherically symmetric systems like the hydrogen atom in its ground state, this constraint substantially reduces the number of independent real-time propagation pathways required to capture the full isotropic response (see further details in Appendix~\ref{app:orient_avg_qp}).

Following an impulsive $\delta$-kick perturbation, the orientation-averaged dipole absorption cross section is given by~\cite{guandalini2021}:
%
%
\begin{equation}\label{eq.sigma_general}
\sigma^{\rm d}_{\rm tot}(\omega) = 
  \frac{4\pi\omega}{3c}  
    \text{Im}\left[\frac{\tilde{d}_i(\omega)\tilde{E}_i(\omega) }{|\widetilde{E}(\omega)|^2} \right],
\end{equation}
where $\tilde{d}(\omega)$ and $\tilde{E}(\omega)$ are the Fourier transforms of the time-dependent dipole moment and the applied field, respectively, and $c$ is the speed of light in vacuum. 
In the presence of a spatially non-uniform field, the total cross section incorporates higher-order multipole contributions.
%
%
In analogy to the dipole case [Eq.~\eqref{eq.sigma_general}], the quadrupolar cross section evaluated at nuclear position $\mathbf{r} = \mathbf{r}_0$ is:
\begin{equation}\label{eq.sigma_quad_general}
\sigma^{\rm qd}_{\rm tot}(\mathbf{r},\omega) = 
  \frac{4\pi\omega}{5c}
    \text{Im}\left[\frac{\widetilde{\Omega}^{\rm qd}_{ij}(\omega)}{|\widetilde{E}(\mathbf{r},\omega)|^2}   \frac{\partial \widetilde{E}_i(\mathbf{r},\omega)}{\partial x_j} \right],
\end{equation}
where the prefactor $1/5$ originates from the orientation averaging of the rank-2 tensor.
The total cross section is expressed as the sum of these response channels:
\begin{equation}\label{eq.sigm_dQ_refined}
\sigma_{\rm tot}(\omega) = \sigma^{\rm d}_{\rm tot}(\omega) + \sigma^{\rm qd}_{\rm tot}(\omega) + \ldots
\end{equation}

All real-time simulations presented hereafter are performed using the \texttt{Octopus} package~\cite{tancogne2020} and validated against a grid-based B-spline basis for the TDSE. Since the hydrogen atom is a one-electron system, the calculations are carried out with the Hartree and exchange-correlation potentials set to zero ($V_{\rm Hxc}=0$). The electron therefore evolves in the bare attractive nuclear potential of Eq.~\eqref{eq:field-free_H}, implying that the description is inherently free of self-interaction errors. In this form, the RT-TDDFT propagation is equivalent to the exact numerical solution of the TDSE. Detailed convergence tests with respect to the simulation box size and the angular momentum expansion are reported in Appendix~\ref{app.tdse_tddft}.

\section{Results and Discussion}\label{sec:Rels}
We apply the theoretical background introduced in Sec.~\ref{sec:theory} to the hydrogen atom, contrasting the numerical results obtained with \texttt{Octopus}~\cite{tancogne2020} against both analytical expressions and numerical solutions of the TDSE for the quadrupole transition matrix elements (Table~\ref{tab:qdt_com}). The \texttt{Octopus} results are in excellent agreement with the reference values, capturing transition amplitudes with a minor deviation of less than 2\%. This small discrepancy is a well-understood consequence of spatial grid discretization, typical for higher-order multipole operators evaluated on a real-space Cartesian mesh, and scales predictably with the grid spacing.
\begin{table}[ht!]
    \centering
    \caption{Absolute values (in atomic units) of the quadrupole transition-matrix elements calculated analytically, via numerical solution of the TDSE, and using \texttt{Octopus}. }
    \begin{tabular}{l c c c}  
        \toprule
        \textbf{Transition} & \textbf{Analytic} & \textbf{TDSE} & \textbf{\texttt{Octopus}} \\  
       \midrule
        $\langle\psi_{1S}|\hat{Q}_{zz}|\psi_{3D_0}\rangle$ & 1.550067 & 1.550068 & 1.575133 \\
       $\langle\psi_{1S}|\hat{Q}_{zz}|\psi_{4D_0}\rangle$ & 1.006632 & 1.006610  & 1.030200  \\
        $\langle\psi_{2S}|\hat{Q}_{zz}|\psi_{3D_0}\rangle$ & 42.105700  & 42.105600  & --       \\
       \bottomrule
    \end{tabular}
    \label{tab:qdt_com}
  \end{table}

\subsection{Absorption Induced by a Spatially Homogeneous Broadband Impulsive Field}\label{ssec.uniform}
 Having validated the accuracy of the numerical implementation, we now investigate the behavior of the photoabsorption cross section across spatially uniform and non-uniform field configurations.
To connect the real-time electronic dynamics to experimental observables, we compute the photo-absorption cross-section [Eq.~\eqref{eq.sigma_general}] for the hydrogen atom under both weak and strong dipolar fields (the WD and SD regimes, respectively, see Table~\ref{tab:regimes}).

\begin{figure}
\centering
\hspace*{-0.5cm}
\includegraphics[width=0.5\textwidth]{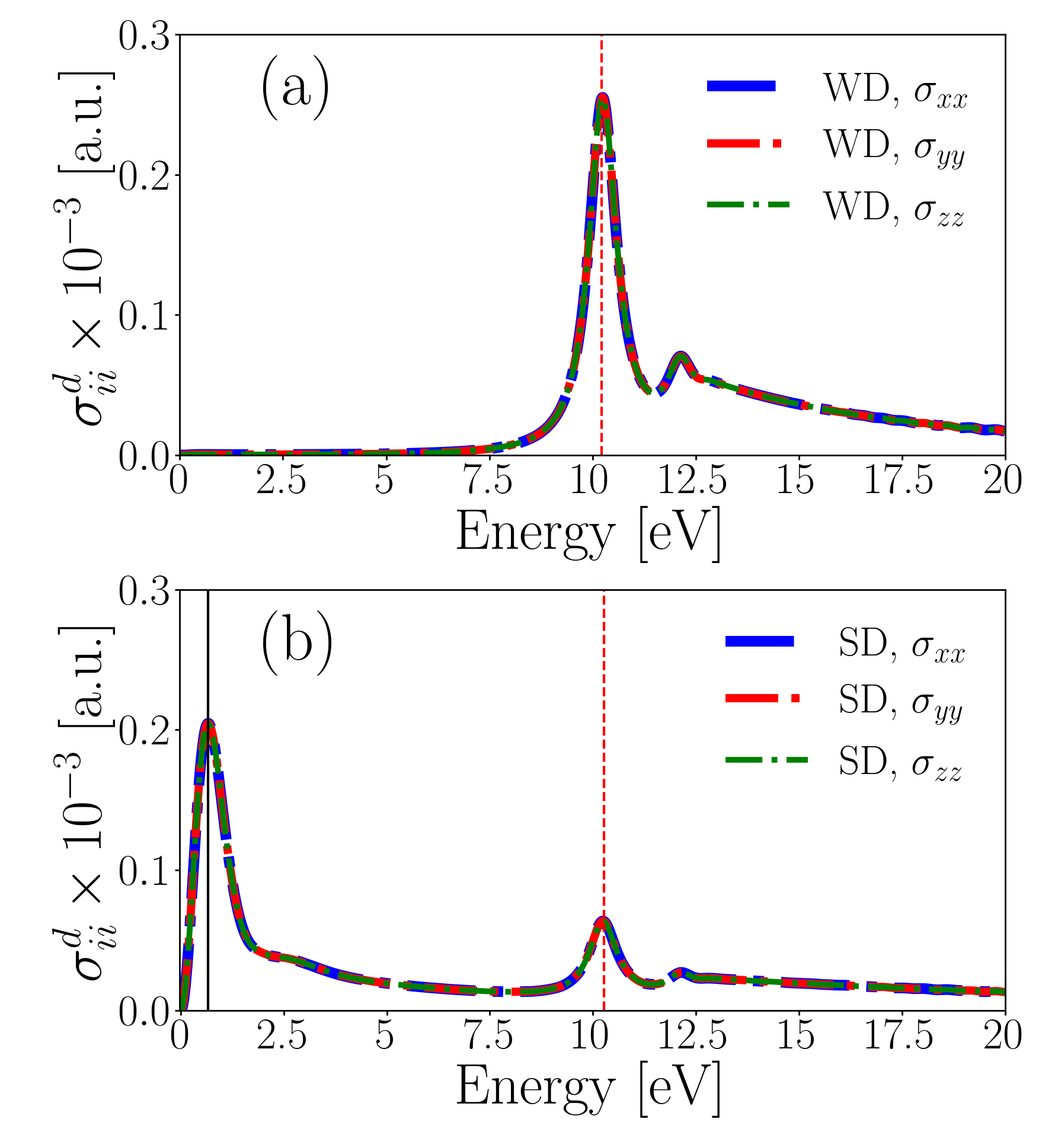}
\caption{Diagonal elements of the photo-absorption cross-section tensor ($\sigma^{d}_{xx}, \sigma^{d}_{yy}, \sigma^{d}_{zz}$) as a function of photon energy. (a) Linear WD regime ($E_0 = 0.05$~a.u.) featuring the strong dipole transition $1s \to 2p$ at 10.22~eV (red dashed bar) and the weaker one $1s \to 3p$ at 12.09~eV. (b) Nonlinear SD regime ($E_0 = 1.0$~a.u.) displaying a severe depletion of the $1s \to 2p$ resonance (red dashed bar) and the emergence of a prominent low-energy peak at $0.60$~eV (solid black bar).}
\label{fig.sigma_lin}
\end{figure}

To explore the linear domain (WD), where the field amplitude satisfies $E_0 \ll E_{\text{th,d}}$ [Eq.~\eqref{eq.dipH_F})], we apply a spatially uniform $\delta$-kick with a magnitude of $E_0 = 0.05$~a.u., corresponding to a peak intensity of $I\approx8.8\cdot10^{13} \; \rm W/cm^2$. Due to the spherical symmetry of the H atom in the ground state,  each diagonal element of the cross-section tensor is identical ($\sigma^{\rm d}_{xx} = \sigma^{\rm d}_{yy} = \sigma^{\rm d}_{zz}$), as confirmed by the perfect overlap of the curves in Fig.~\ref{fig.sigma_lin}a. The calculated linear spectrum exhibits a dominant resonance at 10.22~eV, corresponding to the $1s \to 2p$ dipole transition. A much weaker resonance is resolved around 12.1~eV. Since the $3p$ and $3d$ subshells are energetically degenerate in the hydrogen atom, this resonance cannot be assigned from the excitation energy alone. However, within the linear regime, the transition is dictated by the dipole selection rule ($\Delta \ell = \pm 1$). Starting from the $1s$ ground state ($\ell=0$), the $1s \to 3d$ transition is dipole-forbidden, allowing us to uniquely attribute this linear peak to the $1s \to 3p$ excitation. The alignment of these peaks with the analytical hydrogenic eigenvalues confirms that a weak $\delta$-kick successfully maps the linear-response regime of the atom.

Increasing the field strength to $E_0 = 1$~a.u. drives the system into the nonlinear SD regime ($E_0 \sim E_{\text{th,d}}$), see Fig.~\ref{fig.sigma_lin}b. Here, while the primary peak ($1s \to 2p$ transition) at 10.22~eV remains visible, it undergoes dramatic intensity depletion, decreasing by almost a factor of 5 compared to the WD case. Crucially, the Thomas-Reiche-Kuhn (TRK) sum rule~\cite{TRK1925} remains satisfied: integrating the partial cross-section up to $20$~eV yields $S = 0.591 \times 10^{-3}$~a.u. in the WD regime and $S = 0.588 \times 10^{-3}$~a.u. in the SD regime, corresponding to a variation of about $1\%$. This conservation confirms that the suppression of the main resonance does not stem from a loss of absorption capacity, but rather from a profound redistribution of spectral weight into non-linear channels.

The most striking feature in the SD spectrum is the emergence of a prominent low-energy resonance at 0.60~eV. This intense absorption maximum is characteristic of strong $\delta$-kick excitations~\cite{cocchi2014ab} and can be associated with excited-state absorption pathways recently predicted in oligothiophene molecules~\cite{quad:Driouech2025}. Here, we attribute this feature to transitions between highly excited shells populated during intense $\delta$-kick irradiation. The presence of the strong uniform field drives sequential multiphoton pathways, enabling the system to access high-$\ell$ states that are strictly forbidden via linear single-photon absorption.

To understand how these low-energy channels are accessed under a uniform, instantaneous external field, we examine higher-order terms contributing to the time-evolution operator. Although a spatially homogeneous $\delta$-kick preserves the dipolar form of the interaction Hamiltonian, $\hat{H}_{\rm int}(\mathbf{r},t) = -\mathbf{d}\cdot\mathbf{E}(\mathbf{r}_0,t)$, a strong field of  $E_0 = 1.0$~a.u. imprints a large, non-perturbative phase factor onto the electronic wavefunction, generating finite amplitudes along excitation pathways that are inaccessible in linear response. Consequently, the second-order term of the propagator expansion becomes non-negligible (see Appendix~\ref{app:cros_tdse_tddft} and Table~\ref{tab:regimes}). Since this second-order operator scales quadratically with the spatial electronic coordinates ($\sim x^2$ for an $x$-polarized field), it couples states matching the selection rule $\Delta \ell = 0, \pm 2$. Physically, this describes a sequential two-step dipole process ($1s \to 2p \to 3d$). Thus, even in the absence of a spatial field gradient or explicit quadrupole operator in the Hamiltonian, the intense uniform field drives multiphoton pathways with effective quadrupolar character, populating the high-$\ell$ states responsible for the sub~eV excited-state absorption channels.

\subsection{Absorption Induced by a Spatially Inhomogeneous Broadband Impulsive Field}\label{ssec.Non-uniform}
We continue our analysis by examining the electronic response of the hydrogen atom to a spatially inhomogeneous, broadband impulsive electric field. We consider two distinct configurations: (i) a pure gradient field, which isolates the quadrupole interaction by vanishing at the nuclear position and suppressing the dipole contribution in the current subsection, and (ii) a mixed field in Sec.\ref{ssec.d_QD}, containing both homogeneous and inhomogeneous components, which enables the simultaneous excitation of dipolar and quadrupolar channels.

We begin by discussing the pure quadrupole regime, where the nucleus is located at the origin ($\mathbf{r}_0 = 0$) and the external electric field vanishes as $\mathbf{E}(\mathbf{r}_0,t) = 0$. To isolate the individual quadrupole components, the spatial field configurations are selected to mirror the five real spherical harmonics for $l=2$ [Eqs.~\eqref{eq:real_harmonics_app1}-\eqref{eq:real_harmonics_app5}]. Each scenario is designed to couple exclusively to a single component of the quadrupole moment tensor in the spherical basis. Under these conditions, the dipole interaction is strictly suppressed, allowing the quadrupolar contribution to be isolated via the corresponding photoabsorption cross section $\sigma^{\text{qd}}(\omega)$.

\begin{figure}
\centering
\hspace*{-0.5cm}
\includegraphics[width=0.5\textwidth]{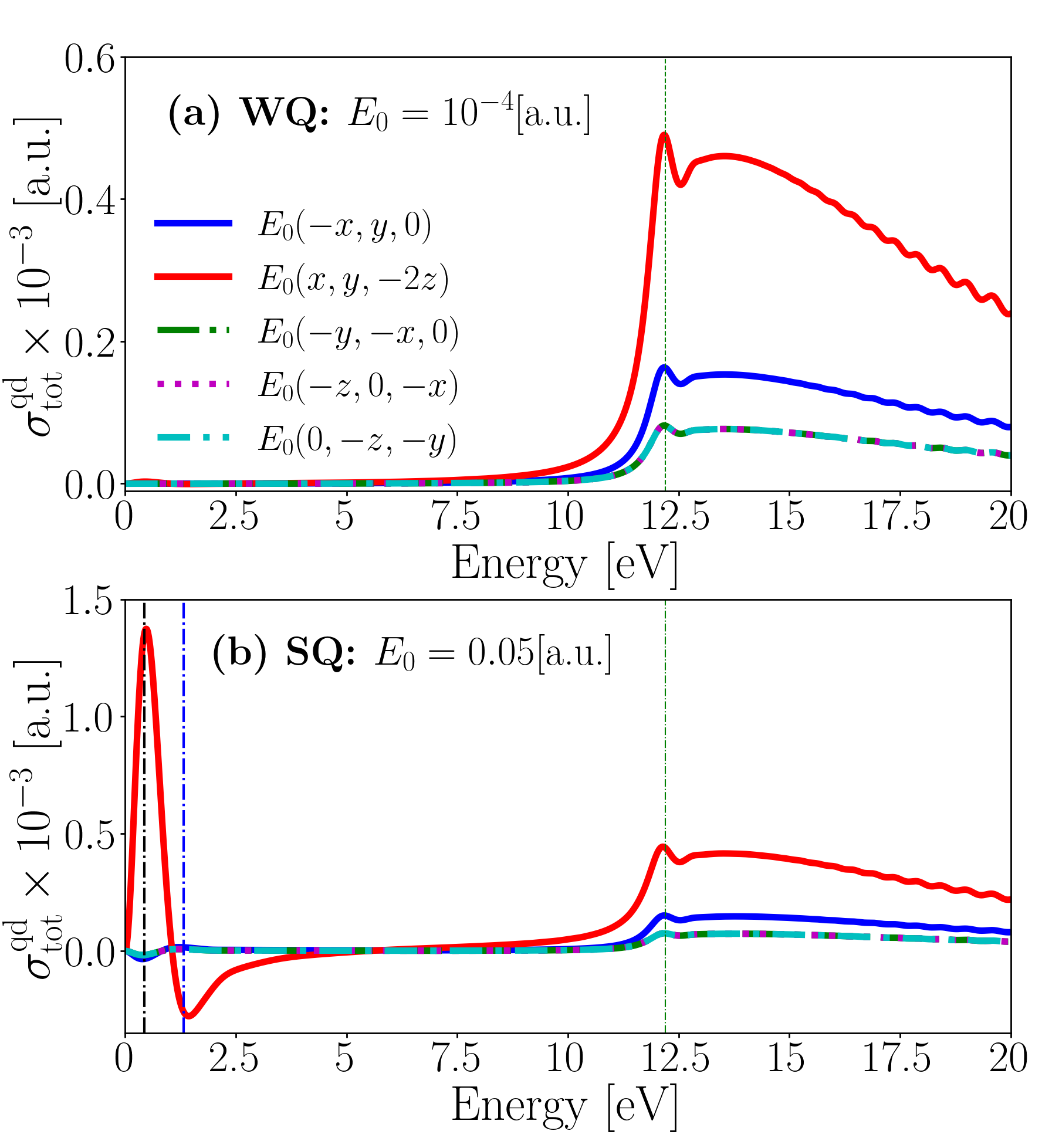}
\caption{Total photo-absorption cross section ($\sigma^{\rm qd}_{\text{tot}}(\omega)$) computed for the hydrogen atom under spatially non-uniform electric field gradients. (a) Linear WQ regime ($E_0 = 10^{-4}$~a.u.) where the three off-diagonal configurations (green dash-dotted, purple dotted, and cyan dash-dot-dotted curves) perfectly superimpose. The solid blue and red curves correspond to the diagonal gradient configurations. The vertical dashed green line highlights the $1s \to 3d$ quadrupole resonance at 12.1~eV. (b) Nonlinear SQ regime ($E_0 = 0.05$~a.u.) illustrating the amplitude splitting among different gradient forms. The vertical black dash-dotted line marks the 0.44~eV induced absorption peak, while the vertical blue dash-dotted line marks the non-linear stimulated emission minimum at 1.32~eV.}
\label{fig.sigma_lin2}
\end{figure}

The WQ regime is accessed with the field amplitude $E_0 = 10^{-4}$~a.u. (gradient scale of $\Gamma_0 = 1$~a.u.), corresponding to a threshold $E_{\rm th,qd} \sim 1/3 \approx 0.33$ a.u. [Eq.~\eqref{eq.thr_FQ1}], ensuring $E_0 \ll E_{\rm th,qd}$ and placing the simulation within the linear regime. For all examined field gradient configurations, the spectrum is characterized by a single, sharp resonance at about 12.1~eV (dashed green bar in Fig.~\ref{fig.sigma_lin2}a). In contrast to the multiphoton absorption features observed under intense uniform fields, this resonance represents a true, single-photon electric-quadrupole transition from the spherically symmetric $1s$ ground state to the degenerate $3d$ manifold, according to the $\Delta \ell = \pm 2$ selection rule. The spatial field gradient provides the necessary spatial phase to drive this transition directly, leading to a substantial enhancement of the absorption cross section at this frequency.

In the SQ regime, the field amplitude is increased to $E_0 = 0.05$~a.u. maintaining $\Gamma_0 = 1$. This setting yields an interaction scale $|H_{qd}| \sim \braket{r^2} E_0 \Gamma_0 \approx 0.15$, placing the system at the onset of the nonlinear crossover. Since the coupling strength is ruled only by the product $E_0 \Gamma_0$, this onset can be reached equivalently by increasing either the amplitude or the gradient (see discussion of Fig.~\ref{fig.sigma_lin_nonli} below), reflecting the gradual, boundary-free nature of the linear-to-nonlinear transition.
The cross sections obtained from the off-diagonal field configurations, evaluated using coordinate-swapped geometries, are perfectly coincident in the WQ regime (green, cyan, and purple curves in Fig.~\ref{fig.sigma_lin2}a). This superposition robustly confirms the rotational invariance of the quadrupole excitation energy under arbitrary rotations of the gradient tensor; a detailed treatment of this invariance is deferred to Appendix~\ref{app:linear_avg}. In the linear regime, the underlying response operator is isotropic and transforms covariantly as a rank-2 tensor satisfying Schur's lemma (see Appendix~\ref{app:sym_simplification} and \ref{app:rotation_Q}). From a practical standpoint, this off-diagonal configuration offers a computationally elegant shortcut: the scalar contraction of the off-diagonal gradient tensor yields the total orientation-averaged cross section directly from a single real-time propagation. Conversely, when employing the field configurations of Eqs.~\eqref{eq:real_harmonics_app3}--\eqref{eq:real_harmonics_app5}, the diagonal elements of the quadrupole tensor are coupled (red and blue curves in Fig.~\ref{fig.sigma_lin2}). 

 In the SQ regime (Fig.~\ref{fig.sigma_lin2}b), the intense field gradient induces a non-linear response where the higher-order contributions transform as higher-rank tensors. Since their projections onto the $\ell = 2$ channel do not commute with standard $SO(3)$ rotations, this anisotropy breaks the degeneracy, resulting in a clear splitting of the cross-section amplitudes among the different diagonal field configurations. Spectroscopically, the SQ regime is characterized by a striking nonlinear signature: the emergence of a minimum at 1.32~eV (blue dashed bar in Fig.~\ref{fig.sigma_lin2}b), accompanied by a pronounced positive peak at 0.44~eV (black dashed-dotted bar). The negative cross-section feature represents a phase-matched stimulated emission channel, where the intense, spatially inhomogeneous kick acts simultaneously as an optical pump and a probe, driving stimulated downward transitions between transiently populated excited states.  These low-energy features correspond to active $\Delta \ell = 0$ quadrupole transitions mediated by the strong gradient operator, analogous to the intensity-induced processes analyzed for the uniform field in Sec.~\ref{ssec.uniform} (further details in Appendix~\ref{app:cros_tdse_tddft}). 
 
 The choice of field geometry dictates how these non-linear spectral signatures can be extracted. In the strong-field domain, the off-diagonal field configurations $\mathbf{E}=\rm E_0(-y, -x, 0)$, $\mathbf{E}= E_0(-z, 0, -x)$, and $\mathbf{E}= E_0(0, -z, -y)$ prove to be highly advantageous. As shown in Fig.~\ref{fig.sigma_lin2}b, they cleanly isolate the sub-eV non-linear signatures and stimulated emission features, while preserving the integrity of the higher-energy maxima. From a practical perspective, because these off-diagonal layouts directly map the isotropic signal without being subject to the diagonal constraints, they offer a computationally elegant route to obtaining the total orientation-averaged cross section from a single reference calculation.

 Conversely, for diagonal configurations [Eqs.~\eqref{eq:real_harmonics_app3}--\eqref{eq:real_harmonics_app5}], the quadrupole tensor elements are linearly dependent and traceless. In the WQ scenario, this constraint is easily handled because the linear response operator is governed by a single scalar frequency coefficient, yielding identical cross-section profiles across the board. However, in the SQ regime, the intense field gradient activates higher-order tensor contributions whose projections onto the $\ell = 2$ channel do not commute with standard $SO(3)$ rotations. This anisotropy lifts the geometric degeneracy, generating the clear splitting in amplitude observed between the red and blue curves in Fig.~\ref{fig.sigma_lin2}b.

To verify the physical consistency and numerical stability of these calculations, we employ the generalized multipole extension of the TRK sum rule, as investigated ín \cite{Wang1999}. As summarized in Table~\ref{tab:qd_TRF}, the TRK sum rule is well preserved across all configurations, converging within 1.5\% of the theoretical limit for the off-diagonal configurations. Even in the intense SQ regime, where the real-space grid dynamics are strongly non-linear, the maximum discrepancy remains under $2.4\%$ for the diagonal $\mathbf{E} = E_0(x,y,-2z)$ profile. This global conservation confirms that the negative absorption peaks in the cross section (Fig.~\ref{fig.sigma_lin2}a) represent a physically robust energy redistribution via phase-matched stimulated emission rather than an artifact of numerical norm degradation. We caution, however, that while the physical origin and sign of these sub-eV features are firmly established, their precise value of their amplitude remains sensitive to boundary effects: because the driving operator ($F = r^2 Y_{2m}$) growsunbounded with distance, the low-energy signal probes the far-field continuum, making its exact magnitude susceptible to absorbing boundary conditions and finite box sizes.

%
%

\begin{table}
    \centering
    \caption{Generalized TRK sum rule for the partial integrated cross section ($S$) across the WQ and SQ regimes for different field-gradient profiles. The values of $S$ are obtained over a window of 40~eV.}
    \begin{tabular}{l c c}  
        \toprule
        Field Configuration & $S_{\rm WQ} \times 10^{-3}$ [a.u.]  & $S_{\rm SQ}\times 10^{-3}$ [a.u.]   \\  
       \midrule
        $E_0(-x,y,0)$ & 1.719675 & 1.719671  \\
       $E_0(x,y,-2z)$ &  5.159278 & 5.284176   \\
        $E_0(-y,-x,0)$ & 0.859743  & 0.859740         \\
        $E_0(-z,0,-x)$ & 0.859743  & 0.859740        \\
        $E_0(0,-z,-y)$ & 0.859743  & 0.859740        \\
       \bottomrule
    \end{tabular}
    \label{tab:qd_TRF}
  \end{table}

To map the exact boundaries of the linear-to-nonlinear transition, we systematically investigate the joint influence of the threshold field strength [Eq.~\eqref{eq.thr_FQ1}] and the spatial field gradient parameters on the resulting spectral features. The linear spatial dependence of the $\delta$-kick perturbation provides an exceptionally flexible framework for isolating the role of the gradient. Specifically, in the absence of a uniform dipole background ($\alpha_i = 0$), the coefficients $\beta_i$, $\gamma_i$, and $\delta_i$ can be modulated independently of the field amplitude $E_0$, unlocking a direct route to study the gradient dependence of the spectrum. Since the cross-section profiles exhibit universal patterns across the different spatial fields (see Fig.~\ref{fig.sigma_lin2}), we focus on a single representative field configuration, namely $\mathbf{E} = E_0(-\beta x, \gamma y, 0)$. To simulate an isotropic expansion where the spatial gradient is enhanced equally along both active Cartesian axes, we set $\beta = \gamma \equiv \Gamma_0$, where $\Gamma_0$ denotes the field gradient strength and is chosen consistently with Eq.~\eqref{eq:alpha_single_calc_app}.  

\begin{figure}
\centering
\includegraphics[width=0.48\textwidth]{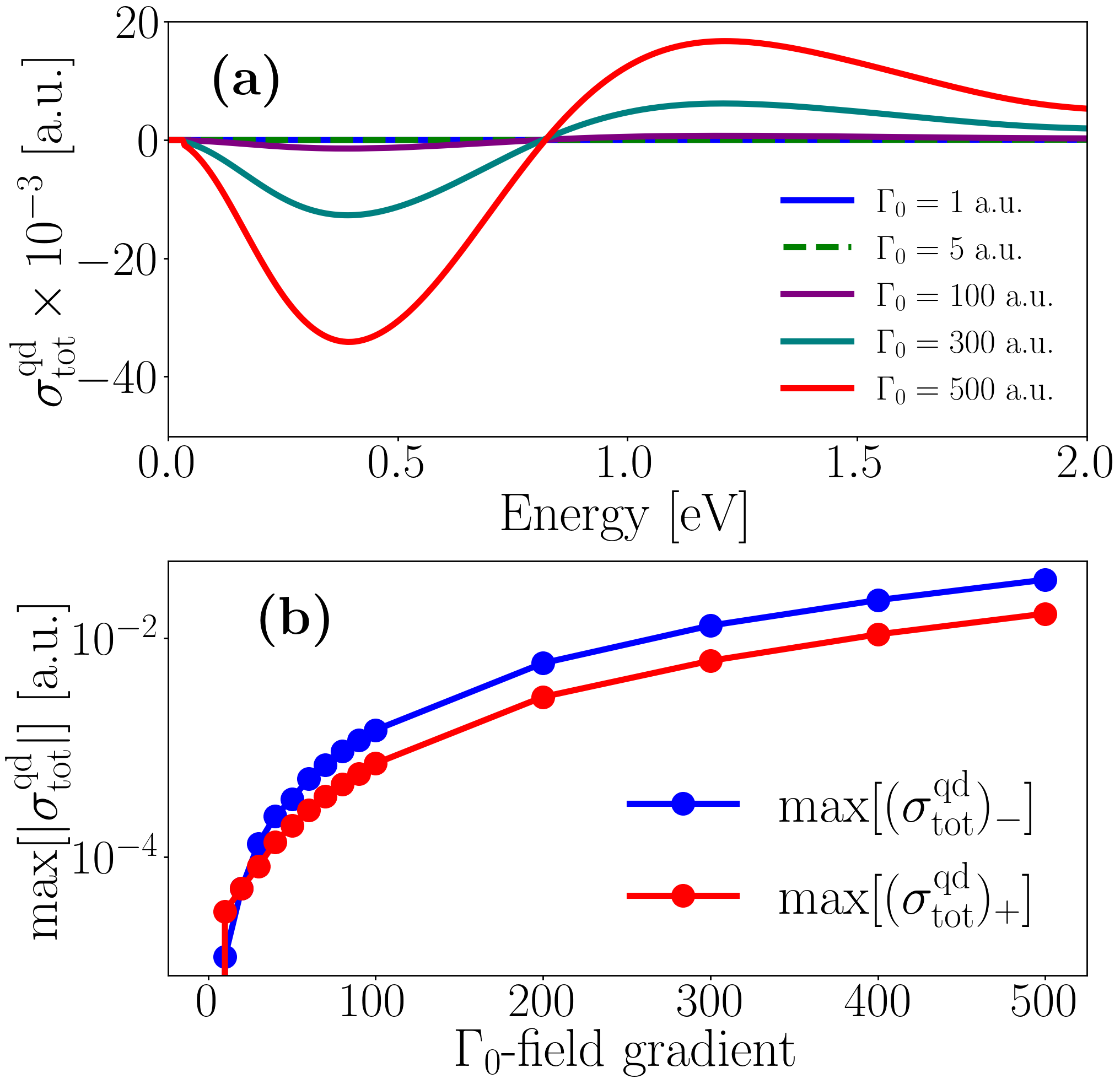} 
\caption{Parametric scaling of the sub-eV nonlinear signatures for the representative field configuration $\mathbf{E} = E_0(-\Gamma_0 x, \Gamma_0 y, 0)$ at a fixed weak amplitude $E_0 = 10^{-4}$~a.u. (a) Quadrupole absorption cross section at varying gradient strengths $\Gamma_0$. (b) Logarithmic plot of the absolute maximum and minimum values of $(\sigma^{qd}_{\rm tot})_{\pm}$ (0.44~eV and 1.32~eV, respectively) as a function of $\Gamma_0$. These extrema are denoted in the legend as $\max[\sigma^{qd}_{+}]$ and $\max[\sigma^{qd}_{-}]$.}
\label{fig.sigma_lin_nonli}
\end{figure}

We adopt a field amplitude of $E_0 = 10^{-4}$~a.u. as a baseline, for which sub-eV spectral features are absent with $\Gamma_0 = 1$. As visualized in Fig.~\ref{fig.sigma_lin_nonli}a, a systematic increase of the field gradient leaves the normalized response essentially unchanged within the weak-to-moderate regime ($\Gamma_0 \le 5$). This invariance confirms that the normalized cross section is independent of $\Gamma_0$ as long as the system remains in the linear regime [Eq.~\eqref{eq:alpha_single_calc_app}]:
\begin{equation}
\begin{aligned}
\Omega : \Gamma_0
&= \sum_{i,j=1}^{3} \Omega_{ij} (\Gamma_0)_{ij}
 = \Gamma_0(-\Omega_{xx}+\Omega_{yy}),\\
\Gamma_0 : \Gamma_0
&= \sum_{i,j=1}^{3} (\Gamma_0)_{ij}(\Gamma_0)_{ij}
 = 2\Gamma_0^2.
\end{aligned}
\label{eq:scalar_prob}
\end{equation}
Here, the first relation evaluates the contraction of the quadrupole response tensor with the gradient profile $\Gamma_0$, while the second yields the squared norm in the cross-section denominator. Since both terms scale identically as $\Gamma_0^2$ in the linear regime, $\Gamma_0$ cancels out, leaving the normalized cross section invariant. Our real-time simulations perfectly reproduce this analytical scaling: for $\Gamma_0 \le 5$, the sub-eV response remains strictly zero. However, when the gradient is pushed beyond this linear threshold ($\Gamma_0 \ge 10$), higher-order non-linear terms take over, driving the gradual emergence of the characteristic low-energy absorption maxima and stimulated emission minima.

To quantify the onset of the nonlinear behavior, we display on a logarithmic scale the absolute peak values of the low-energy peaks as a function of $\Gamma_0$ (Fig.~\ref{fig.sigma_lin_nonli}b). Even at the highest simulated gradients, these induced peak values remain 2–3 orders of magnitude smaller than the corresponding quadrupole transition strength (vertical dashed red line in Fig.~\ref{fig.sigma_lin2}). This result confirms that, despite the opening of nonlinear channels at low energies, the interaction remains predominantly linear in the explored range of $\Gamma_0$. Furthermore, because the non-perturbative coupling depends on the product $E_0 \Gamma_0$, this gradient-induced scaling trend can be directly mapped to the case of holding the gradient fixed at $\Gamma_0 = 1$~a.u. while systematically elevating the field amplitude $E_0$. This equivalence reinforces our key finding: once the cumulative interaction parameters overcome the characteristic threshold value, the electronic wavepacket undergoes a continuous transition into the nonlinear regime. Since this linear-to-nonlinear crossover lacks a sharp boundary, the analytical threshold expression in Eq.~\eqref{eq.thr_FQ1} offers a useful qualitative guide for locating the onset of non-linear quadrupolar light-matter interactions. The gradients scanned here at $E_0 = 10^{-4}$~a.u. reach products $E_0\Gamma_0 \lesssim 10^{-2}$~a.u., still well below the value $E_0\Gamma_0 \approx 0.05$~a.u. ($|H_{qd}| \approx 0.15$) that characterizes the SQ regime of Fig.~\ref{fig.sigma_lin2}b; this is precisely why the gradient-induced features that emerge for $\Gamma_0 \gtrsim 10$ remain $2$--$3$ orders of magnitude below the main quadrupole line—they are detectable precursors of, rather than the fully developed, nonlinear response, whose spectroscopic prominence sets in only as the product $E_0\Gamma_0$ approaches the $ \sim 0.1$ crossover.

\subsection{Absorption Induced by a Spatially Inhomogeneous Broadband Impulsive Field: Dipole and Quadrupole Contributions}\label{ssec.d_QD}

To complete our analysis, we finally examine the general scenario in which the external perturbation incorporates both spatially uniform and non-uniform components, thereby driving simultaneous dipolar and quadrupolar electronic transitions. This mixed configuration is realized by superimposing a uniform field onto the inhomogeneous gradient profile. For instance, building on the pure gradient configuration $\mathbf{E} = E_0(-x, 0, -z)$, we add a uniform component along $y$, yielding the mixed layout $\mathbf{E} = E_0(-x, 1, -z)$.

\begin{figure}[ht]
\centering
\includegraphics[width=0.5\textwidth]{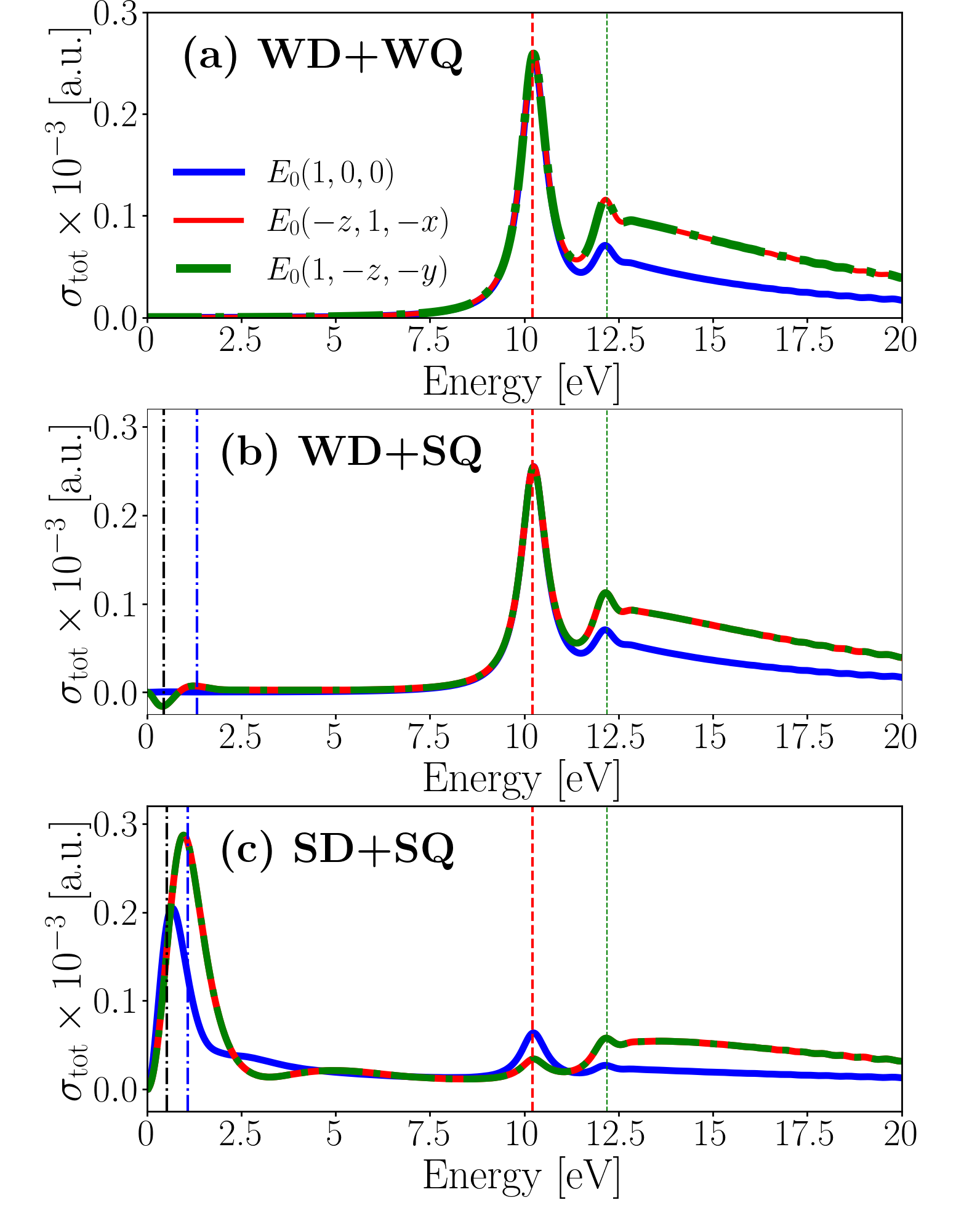}
\caption{Total photo-absorption cross section of the hydrogen atom subject to mixed spatial perturbations combining both uniform and non-uniform field dependencies. (a) Weak-field mixed regime (WD+WQ, $E_0 = 10^{-4}$~a.u.). (b) intermediate mixed regime (WD+SQ, $E_0 = 0.05$~a.u.). (c) Strong mixed regime (SD+SQ, $E_0 = 1$~a.u.). The red bar in all panels highlights the $1s \to 3d$ resonance at 12.1~eV, while the black and blue vertical dash-dotted lines in panels (b) and (c) mark the sub-eV non-linear spectral shift relative to the low-energy domain.}
\label{fig.sigma_lin3}
\end{figure}

The characteristic scaling behavior is governed by the interaction thresholds established in Sec.~\ref{sec:theory}: the spatially uniform dipole channel is activated near $E_{\rm th,d} \sim 1$ a.u. (Sec.~\ref{ssec.uniform}), the pure quadrupole threshold sits at $E_{\rm th,qd} \sim 1/(\braket{r^2}\Gamma_0) \approx 0.3$~a.u. with $\Gamma_0 = 1$ (Sec.~\ref{ssec.Non-uniform}), and the mixed-field configuration threshold is given by $E_{\rm th,d+qd} \sim 1/(1+3\Gamma_0) \approx 0.25$~a.u. at $\Gamma_0 = 1$ [Eq.~\eqref{eq.thr_FQ}]. Since the applied amplitude $E_0 = 10^{-4}$ a.u. lies several orders of magnitude below all these box-independent thresholds, the interaction remains placed firmly in the weak-interaction regime (Fig.~\ref{fig.sigma_lin3}a).
Here, dipolar and quadrupolar channels superimpose linearly without mutual cross-channel interference: the primary dipole peak at $10.22$~eV remains fixed and insensitive to spatial gradients, while the $1s \to 3d$ quadrupole resonance at $12.1$~eV (vertical dashed green line) exhibits only a modest enhancement relative to the purely dipolar spectrum (Fig.~\ref{fig.sigma_lin2}a).

Upon increasing the pulse strength to $E_{\rm 0} = 0.05$~a.u., the system selectively enters the nonlinear regime via its quadrupolar response (Fig.~\ref{fig.sigma_lin3}b). As the field strength satisfies $E_{\rm th,d+qd} < E_{\rm 0} \ll E_{\rm th,d}$, the dipole channel remains linear while the quadrupole interaction crosses its non-linear threshold. Consequently, while the $10.22$~eV dipole lines remain perfectly coincident, the sub-eV domain exhibits the characteristic non-linear signatures observed in the pure quadrupole case, specifically, the induced absorption $1.25$~eV and the phase-matched stimulated emission minimum near $0.44$~eV. The overlap of the rotated field configurations in this panel demonstrates that structural rotational symmetry remains robust at the onset of quadrupolar nonlinearity. Simulations using alternative mixed-field configurations, such as $\mathbf{E} = E_0(1, -z, -y)$, overlap perfectly across both Figs.~\ref{fig.sigma_lin3}a and \ref{fig.sigma_lin3}b, confirming that global rotational covariance remains robust from the linear regime up to the onset of quadrupolar non-linearity.

Finally, in the strong-field regime ($E_0 = 1.0$~a.u.) where the field amplitude meets or exceeds all physical thresholds (Fig.~\ref{fig.sigma_lin3}c), the independent multipole channel picture completely breaks down. The primary dipole resonance at $10.22$~eV is severely suppressed and no longer coincides across different spatial configurations, featuring a substantial, anisotropic redistribution of spectral weight. Concurrently, the sub-eV non-linear features are strongly amplified, with the quadrupolar line intensity ultimately surpassing the depleted dipole peak. These profile shifts between non-linear configurations (black vs. blue vertical bars in Fig.~\ref{fig.sigma_lin3}c) confirm that under intense, spatially non-uniform fields, electronic dynamics cannot be modeled via isolated multipole channels, but require a fully coupled, non-perturbative treatment of all active interaction channels.



\section{Summary and Conclusions}
In summary, we have systematically investigated the quadrupole interaction channel in the absorption cross section of the hydrogen atom, serving as the prototypical benchmark to explore the influence of spatial electric field gradients. The structural simplicity of this single-electron system enabled a rigorous analytical and numerical validation, via the numerical solutions of the TDSE contrasted against our implementation in the RT-TDDFT code \texttt{Octopus}. The excellent agreement achieved across these methods establishes the reliability of our development.

Our findings resolve several critical physical mechanisms that govern strong-field electronic dynamics under spatially uniform and non-uniform instantaneous perturbations. In the weak-field domain, the spatial field gradient selectively drives the single-photon $1s \to 3d$ quadrupole resonance at 12.1~eV, while the dipole channel remains dominated by the fundamental $1s \to 2p$ transition at 10.22~eV. Across this linear regime, the response operator remains isotropic, and evaluating the quadrupole contribution through coordinate-swapped off-diagonal configurations preserves rotational covariance. In the strong-field domain, distinct sub-eV excited-state pathways are activated: uniform fields drive multi-step dipole sequences depleting the primary resonance, whereas pure gradient fields yield a stimulated emission minimum near 1.32~eV alongside a positive absorption maximum around 0.44~eV. While intense gradients induce amplitude splitting among diagonal tensor configurations due to non-commuting SO(3) projections, the off-diagonal layout bypasses these geometric constraints, yielding the orientation-averaged cross section. Finally, under fully mixed fields, the independent multipole picture breaks down, giving rise to strong dipole-quadrupole coupling, massive oscillator strength redistribution, and pronounced sub-eV spectral shifts. 

In conclusion, this work establishes a rigorous foundation for simulating near-field excitations driven by localized electromagnetic field gradients across realistic molecular architectures and nanostructured environments, offering crucial theoretical capabilities for the advancing frontiers of near-field nano-optics, structured-light spectroscopy, and spatial-gradient attosecond science.




\section*{Acknowledgments}
This work was supported by the German Research Foundation, project numbers 398816777 (CRC 1375, sub-project A8) and 524452181 (INPULS). Computational resources were provided by the National High Performance Computing Alliance (NHR), project ID number nip00074.

\section*{Author Contributions}
\noindent
\textbf{Anvar Khujakulov}: Data Curation (lead); Investigation (lead); Methodology (equal); Software (lead); Visualization (lead); Writing – Original Draft (lead). 
\textbf{Michele Guerrini}: Methodology (equal); Validation (equal); Writing – Review \& Editing (supporting). 
\textbf{Carlo A. Rozzi}: Conceptualization (equal); Methodology (supporting); Validation (equal); Writing – Review \& Editing (equal). 
\textbf{Caterina Cocchi} Conceptualization (equal); Funding Acquisition (lead); Project administration (lead); Supervision (lead); Writing – Review \& Editing (equal).

\section*{Conflict of Interest}
The authors have no conflicts to disclose.

\section*{Data Availability}
The data that support the findings of this study are available in the public repository Zenodo at the following DOI:10.5281/zenodo.21471798.

\appendix
\section{Rotational invariance of quadrupole absorption}
\label{app:orient_avg_qp}

In this Appendix, we summarize the tensor relations needed to interpret the quadrupole contribution to the absorption signal. The quadrupole operator couples to the symmetric traceless part of the electric-field gradient tensor, which transforms as the \(\ell=2\) irreducible representation of \(\mathrm{SO}(3)\)~\cite{Varshalovich1988,Zare1988}. We first recall the linear orientational average in this five-dimensional subspace and then show how the expression simplifies for spherically symmetric systems such as the hydrogen atom considered in this work.

\subsection{Algebraic setting}
\label{app:orient_setting}

Let \(V_2\) denote the real vector space of \(3\times3\) symmetric traceless Cartesian tensors,
\begin{equation}
  V_2=\{T\in\mathbb{R}^{3\times3}\;|\;T_{ij}=T_{ji},\ \mathrm{Tr}\,T=0\},
\end{equation}
with \(\dim V_2=5\), which carries the irreducible $\ell=2$ representation of \(\mathrm{SO}(3)\) under the transformation:
\begin{equation}
  T \mapsto R T R^T.
\end{equation}
We equip $V_2$ with the standard Frobenius inner product,
\begin{equation}
  A{:}B=\mathrm{Tr}(A^TB)=\sum_{ij}A_{ij}B_{ij}.
  \label{eq:frob_app}
\end{equation}

The quadrupole perturbation couples to the symmetric traceless field-gradient tensor
\begin{equation}
  \Gamma_{ij}
  =
  \frac{1}{2}\bigl(\partial_i E_j+\partial_j E_i\bigr)
  -\frac{1}{3}\delta_{ij}\,\nabla\!\cdot\!\mathbf{E}.
  \label{eq:Gamma_def_app}
\end{equation}
For perturbations derived from scalar potentials satisfying Laplace's equation, $\nabla^2\varphi=0$, the field is source-free ($\nabla\!\cdot\!\mathbf{E} = -\nabla^2\varphi = 0$), ensuring $\bm{\Gamma} \in V_2$.

The induced quadrupole moment $\Omega_{ij}(\omega)$, defined as the Fourier transform of the expectation value of the operator $\Omega_{ij}(t)$ in Eq.~\eqref{eq.OmeQ} [see $\hat Q_{ij}$ in Eq.~\eqref{eq.quadrO}], is likewise a symmetric traceless Cartesian tensor belonging to $V_2$.
In the complex spherical basis, the Cartesian tensor components are mapped to rank-2 spherical components as:
\begin{align}
  T^{(2)}_{\pm2} &= \tfrac12(T_{xx}-T_{yy}) \pm i\,T_{xy},
  \label{eq:sph_pm2_app}\\
  T^{(2)}_{\pm1} &= \mp \tfrac{1}{\sqrt2}(T_{xz}\pm i\,T_{yz}),
  \label{eq:sph_pm1_app}\\
  T^{(2)}_0 &= \tfrac{1}{\sqrt6}(2T_{zz}-T_{xx}-T_{yy}).
  \label{eq:sph_0_app}
\end{align}
The Frobenius product then corresponds to the scalar invariant:
\begin{equation}
  A{:}B
  =
  \sum_{m=-2}^{2}
  (-1)^m A^{(2)}_m B^{(2)}_{-m}.
  \label{eq:frob_sph_app}
\end{equation}

The real spherical harmonics used by \texttt{Octopus} relate to the complex spherical harmonics through:
\begin{align}
Y^{\mathrm{real}}_{l,m}
&=
\frac{(-1)^m}{\sqrt{2}}
\left(Y_{lm}+Y_{lm}^{*}\right),
\qquad m>0,
\\
Y^{\mathrm{real}}_{l,0}
&=
Y_{l0},
\\
Y^{\mathrm{real}}_{l,-m}
&=
\frac{i}{\sqrt{2}}
\left(Y_{lm}-Y_{lm}^{*}\right),
\qquad m>0.
\end{align}
For $l=2$, this yields:
\begin{align}
r^2Y^{\mathrm{real}}_{2,-2}
&=
\sqrt{\frac{15}{4\pi}}\,xy,
\label{eq:real_harmonics_app1}
\\
r^2Y^{\mathrm{real}}_{2,-1}
\label{eq:real_harmonics_app2}
&=
\sqrt{\frac{15}{4\pi}}\,yz,
\\
r^2Y^{\mathrm{real}}_{2,0}
&=
\sqrt{\frac{5}{16\pi}}
(2z^2-x^2-y^2),
\label{eq:real_harmonics_app3}
\\
r^2Y^{\mathrm{real}}_{2,1}
&=
\sqrt{\frac{15}{4\pi}}\,xz,
\label{eq:real_harmonics_app4}
\\
r^2Y^{\mathrm{real}}_{2,2}
&=
\sqrt{\frac{15}{16\pi}}
(x^2-y^2).
\label{eq:real_harmonics_app5}
\end{align}
Denoting the quadrupole moments computed by \texttt{Octopus}as $o_m = \langle r^2 Y^{\mathrm{real}}_{2m} \rangle$, the Cartesian traceless quadrupole tensor is reconstructed as:
\begin{subequations}
\label{eq:car_spher_app}
\begin{align}
Q_{zz}
&=
\sqrt{\frac{16\pi}{5}}\,o_0,
\\[4pt]
Q_{xx}
&=
-\frac12 Q_{zz}
+
\frac32\sqrt{\frac{16\pi}{15}}\,o_2,
\\[4pt]
Q_{yy}
&=
-\frac12 Q_{zz}
-
\frac32\sqrt{\frac{16\pi}{15}}\,o_2,
\\[4pt]
Q_{xy}
&=
3\sqrt{\frac{4\pi}{15}}\,o_{-2},
\\[4pt]
Q_{xz}
&=
3\sqrt{\frac{4\pi}{15}}\,o_1,
\\[4pt]
Q_{yz}
&=
3\sqrt{\frac{4\pi}{15}}\,o_{-1}.
\end{align}
\end{subequations}

\subsection{Linear orientational average}
\label{app:linear_avg}

In the linear regime, the quadrupole moment in the spherical basis is given by:
\begin{equation}
  \Omega^{(2)}_m(\omega)
  =
  \sum_{m'=-2}^{2}
  \alpha^{(2)}_{mm'}(\omega)\,
  \Gamma^{(2)}_{m'}(\omega),
  \label{eq:linear_response_app}
\end{equation}
where $\alpha^{(2)}(\omega)$ is a $5\times5$ response operator acting on $V_2$.

Under an arbitrary rotation $R\in\mathrm{SO}(3)$, rank-2 spherical tensors transform as:
\begin{equation}
  T^{(2)}_m
  \mapsto
  \sum_n D^{(2)}_{mn}(R)\,T^{(2)}_n,
  \label{eq:rank2_rot_app}
\end{equation}
and the response operator as:
\begin{equation}
  \alpha^{(2)}_{mm'}(R)
  =
  \sum_{nn'}
  D^{(2)}_{mn}(R)\,
  \alpha^{(2)}_{nn'}\,
  D^{(2)\,*}_{m'n'}(R).
  \label{eq:alpha_rot_app}
\end{equation}

The orientational average over the rotation group $\mathrm{SO}(3)$ is defined by
\begin{equation}
  \left\langle \alpha^{(2)}_{mm'} \right\rangle_{\mathrm{or}}
  =
  \int_{\mathrm{SO}(3)} dR\;
  \alpha^{(2)}_{mm'}(R),
  \label{eq:or_avg_def_app}
\end{equation}
where $dR$ is the normalized Haar measure ($\int_{\mathrm{SO}(3)} dR = 1$). Using Wigner $D$-matrix orthogonality,
\begin{equation}
  \int_{\mathrm{SO}(3)} dR\;
  D^{(j)}_{mn}(R)D^{(j')\,*}_{m'n'}(R)
  =
  \frac{1}{2j+1}
  \delta_{jj'}\delta_{mm'}\delta_{nn'},
  \label{eq:D_orth_app}
\end{equation}
one finds that the averaged response operator simplifies to:
\begin{equation}
  \left\langle \alpha^{(2)}_{mm'} \right\rangle_{\mathrm{or}}
  =
  \frac{\delta_{mm'}}{5}
  \sum_{n=-2}^{2}\alpha^{(2)}_{nn}(\omega).
\end{equation}
Thus, by Schur's lemma, the orientationally averaged response scalar $\bar{\alpha}^{(2)}(\omega)$ reduces to the normalized trace over $V_2$:
\begin{equation}
   \bar{\alpha}^{(2)}(\omega)
  =
  \frac{1}{5}
  \mathrm{Tr}_{V_2}\,\alpha^{(2)}(\omega)
  =
  \frac{1}{5}\sum_{m=-2}^{2}\alpha^{(2)}_{mm}(\omega),
  \label{eq:alpha_bar_app}
\end{equation}
and the orientationally averaged quadrupole absorption cross sectionbecomes:
\begin{equation}
  \left\langle \sigma^{(1)}_{\mathrm{qp}}(\omega)\right\rangle_{\mathrm{or}}
  \propto
  \omega\,
  \mathrm{Im}\!\left[\bar{\alpha}^{(2)}(\omega)\right]\,
  \Gamma(\omega){:}\Gamma^*(\omega).
  \label{eq:sigma_linear_app}
\end{equation}
Here the $\mathrm{SO}(3)$ average replaces the explicit tensor contraction $\widetilde\Omega^{\rm qd}_{ij}(\omega)\,\partial_jE_i(\omega)$ by the scalar product $\bar{\alpha}^{(2)}(\omega)\,\Gamma(\omega){:}\Gamma^*(\omega)$,
identifying $\sigma^{\rm qd}_{\rm av}(\omega)$ in
Eq.~\eqref{eq.sigm_dQ_refined} with
$\langle\sigma^{(1)}_{\mathrm{qp}}(\omega)\rangle_{\mathrm{or}}$.

For a general non-spherical system, evaluating $\mathrm{Tr}_{V_2}\alpha^{(2)}$ requires reconstructing the response in the five-dimensional space $V_2$, which can be achieved using five linearly independent field-gradient configurations. For a spherically symmetric system, however, this
procedure simplifies further.

\subsection{Spherically symmetric systems}
\label{app:sym_simplification}

For systems invariant under the full SO(3) rotation group, such as atoms in an $s$-type ground state, the response operator commutes with all rotation operators. By Schur's lemma, the polarizability tensor is isotropic prior to any orientational averaging,
\begin{equation}
  \alpha^{(2)}_{mm'}(\omega)
  =
  \bar{\alpha}^{(2)}(\omega)\delta_{mm'}.
  \label{eq:alpha_iso_app}
\end{equation}
Equivalently, in Cartesian tensor notation,
\begin{equation}
  \Omega_{ij}(\omega)
  =
  \bar{\alpha}^{(2)}(\omega)\,
  \Gamma_{ij}(\omega).
  \label{eq:Q_alphaG_iso_app}
\end{equation}

The scalar polarizability response $\bar{\alpha}^{(2)}(\omega)$ can therefore be extracted from a single calculation using
any non-vanishing traceless gradient profile $\Gamma_0\in V_2$:
\begin{equation}
  \bar{\alpha}^{(2)}(\omega)
  =
  \frac{\Omega(\omega){:}\Gamma_0}
       {\Gamma_0{:}\Gamma_0}.
  \label{eq:alpha_single_calc_app}
\end{equation}
Thus, even though an arbitrary non-spherical molecule requires five independent gradient fields to reconstruct the trace over $V_2$, a spherically symmetric atom requires only one nonzero quadrupolar perturbation. Rotating the field gradient merely changes the spatial representation of the same underlying scalar response.

Under a spatial rotation $R \in \mathrm{SO}(3)$, the field-gradient tensor and the induced quadrupole tensor transform as
\begin{equation}
  \Gamma^{(R)} = R\Gamma R^T,
  \qquad
  \Omega^{(R)} = R \Omega R^T .
  \label{eq:Gamma_Q_rotation_app}
\end{equation}
For a spherically symmetric system, Eq.~\eqref{eq:Q_alphaG_iso_app} is covariant under this transformation:
\begin{equation}
  \Omega^{(R)}(\omega)
  =
  R\Omega(\omega)R^T
  =
  \bar{\alpha}^{(2)}(\omega)\,
  R\Gamma(\omega)R^T
  =
  \bar{\alpha}^{(2)}(\omega)\Gamma^{(R)}(\omega).
  \label{eq:covariant_response_app}
\end{equation}

Consequently, while different orientations of the quadrupole perturbation may excite different Cartesian tensor components, or different linear combinations of the degenerate $d$-like excited states, they leave the physical absorption spectrum strictly invariant. Any apparent discrepancies between spectra extracted from individual Cartesian components are purely tensor-projection effects, rather than a breakdown of fundamental rotational invariance.

\subsection{Transformation of quadrupole components under spatial rotations}
\label{app:rotation_Q}
\begin{figure*}
\centering
\includegraphics[width=0.95\textwidth]{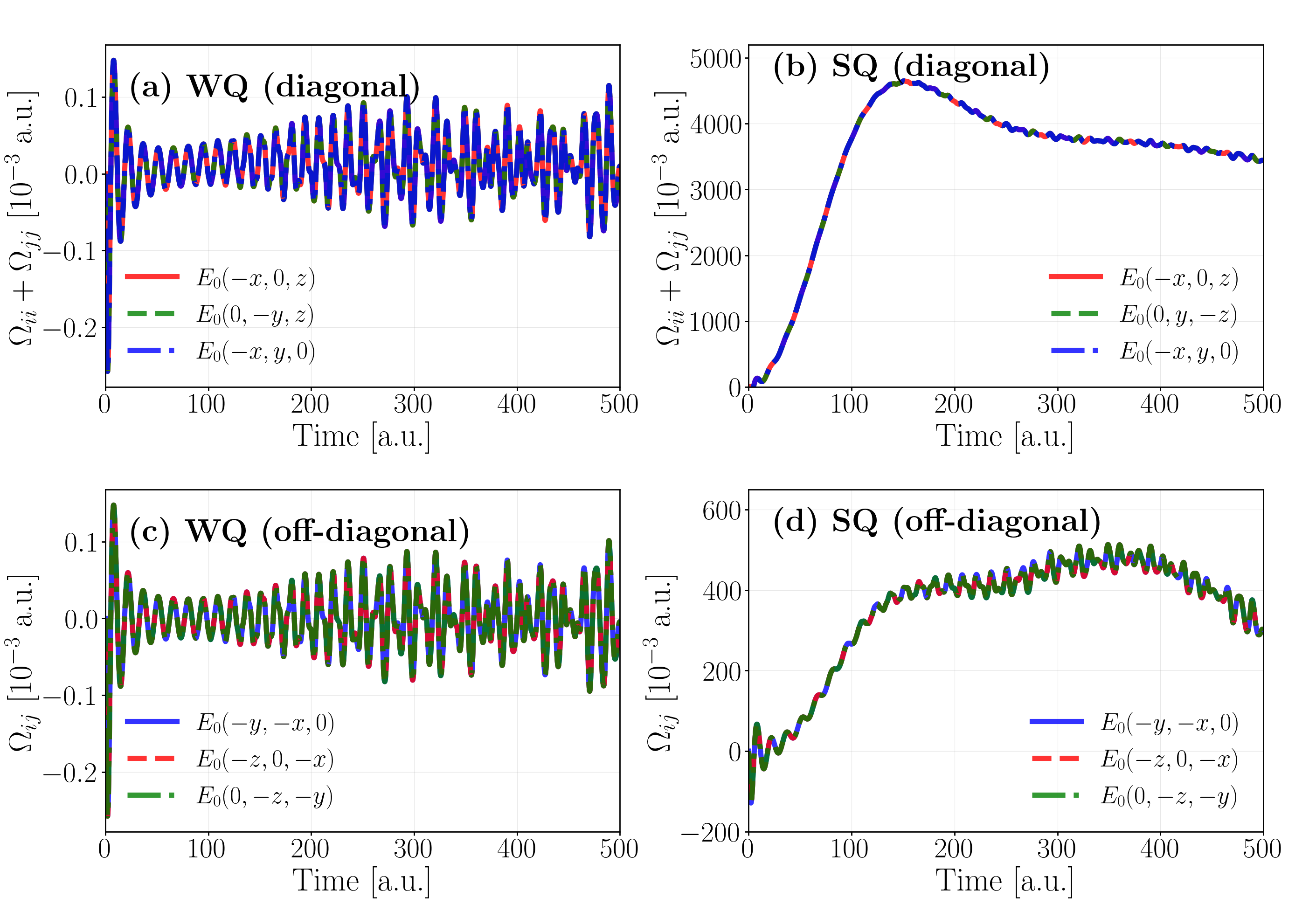}
\caption{Time-dependent diagonal and off-diagonal elements of the quadrupole moment (a-c) in the WQ ($E_0 = 10^{-4}$~a.u.) and (b-d) SQ ($E_0 = 0.05$~a.u.) regimes under the specified non-uniform field dependencies.}
\label{fig.sigma_appQ}
\end{figure*}

The time-dependent quadrupole moments associated with the diagonal
and off-diagonal field-gradient configurations are shown in
Fig.~\ref{fig.sigma_appQ}.  Applying a passive rotation $R_x(\pi/2)$ about the $x$ axis transforms the diagonal gradient
$\Gamma=\operatorname{diag}(-1,1,0)$ into
$\Gamma'=\operatorname{diag}(-1,0,1)$.  The Cartesian quadrupole tensor transforms accordingly under $R_x(\pi/2) \bm{\Omega} R_x^T(\pi/2)$:
\begin{equation}
\begin{pmatrix}
\Omega_{xx} & \Omega_{xy} & \Omega_{xz} \\
\Omega_{yx} & \Omega_{yy} & \Omega_{yz} \\
\Omega_{zx} & \Omega_{zy} & \Omega_{zz}
\end{pmatrix}
\xrightarrow{R_x(\pi/2)}
\begin{pmatrix}
\Omega_{xx} & \Omega_{xz} & -\Omega_{xy} \\
\Omega_{zx} & \Omega_{zz} & -\Omega_{zy} \\
-\Omega_{yx} & -\Omega_{yz} & \Omega_{yy}
\end{pmatrix}.
\label{eq:Q_rot_matrix_app}
\end{equation}

In the diagonal sector, this rotation permutes the diagonal elements according to:
\begin{equation}
\begin{pmatrix}
\Omega_{xx}\\
\Omega_{yy}\\
\Omega_{zz}
\end{pmatrix}
\xrightarrow{R_x(\pi/2)}
\begin{pmatrix}
\Omega_{xx}\\
\Omega_{zz}\\
\Omega_{yy}
\end{pmatrix}.
\label{eq:Q_rot_diag_app}
\end{equation}
Because the tensor is strictly traceless ($\Omega_{xx}+\Omega_{yy}+\Omega_{zz}=0$) the diagonal components must be treated collectively.  The trace-free constraints for the three diagonal field-gradient configurations are:
\begin{align}
\Omega^{(1)}_{xx}+\Omega^{(1)}_{zz} &= -\Omega^{(1)}_{yy}
\quad \text{for } \mathbf E=E_0(-x,y,0), \\
\Omega^{(2)}_{xx}+\Omega^{(2)}_{zz} &= -\Omega^{(2)}_{yy}
\quad \text{for } \mathbf E=E_0(-x,0,z), \\
\Omega^{(3)}_{yy}+\Omega^{(3)}_{zz} &= -\Omega^{(3)}_{xx}
\quad \text{for } \mathbf E=E_0(0,-y,z).
\end{align}
The off-diagonal field configurations bypass these diagonal sum constraints. 
For example, the same $R_x(\pi/2)$ rotation maps $\mathbf E=E_0(-y,-x,0)$ into $\mathbf E=E_0(-z,0,-x)$.  Although individual off-diagonal tensor elements may undergo sign inversion under rotation, the scalar contraction $\Omega{:}\Gamma$ remains invariant.

\section{Convergence and Comparison of \texttt{Octopus} and TDSE Results}\label{app.tdse_tddft}
In this Appendix, we validate the numerical convergence and accuracy of the real-space grid implementation in \texttt{Octopus} of dipole and quadrupole observables through a benchmark with the exact numerical solution of the TDSE. The numerical simulation parameters for both solvers are summarized in Table~\ref{tab:numerical_params}.

\begin{table*}
\centering
\small
\caption{Numerical settings (in a.u.) for the TDSE and \texttt{Octopus} calculations.}
\label{tab:numerical_params}
\begin{tabular}{@{} lcc @{}}
\toprule
\textbf{Parameter} & \textbf{TDSE} & \textbf{\texttt{Octopus}} \\ \midrule
Box radius ($R_{\text{max}}$) & 100 & 100 \\
Propagation Time ($T$) & 1000 & 1000 \\
Time Step ($\Delta t$) & 0.01 & 0.05 \\
Radial Basis / Grid & 200 B-splines (order 9) & $\Delta r = 0.2$ \\
Boundary Condition & $P(R_{\text{max}})=0$ & None \\
Max Angular Momentum & $\ell_{\text{max}}=10$ & \\
\bottomrule
\end{tabular}
\end{table*}
\subsection{Time-dependent dipole and quadrupole from the TDSE and \texttt{Octopus}}\label{ap:tdse_dft}

All real-time simulations were performed using an impulsive $\delta$-kick excitation. The numerical parameters for both the TDSE (B-spline basis) and the RT-TDDFT solver implemented in \texttt{Octopus} are summarized in Table~\ref{tab:numerical_params}. To optimally balance accuracy and computational costs, TDSE results were obtained using $\ell = 10$, having checked convergence by increasing $\ell_{\text{max}}$ up to 14. 

To verify the numerical stability of the ground state, we calculated the total electronic energy of the hydrogen atom using both frameworks (TDSE and \texttt{Octopus}), contrasting the numerical results against the exact reference. The results, reported in Table~\ref{tab:E_tot}, reveal perfect agreement between the TDSE and the exact results. On the other hand, the sub-mHa discrepancies in the \texttt{Octopus} results are attributed to the discretization of the Coulomb potential on a finite real-space grid, a well-documented behavior in grid-based codes \cite{marques2003,castro2006,andrade2015,tancogne2020}.
\begin{table}[h!]
    \centering
    \begin{tabular}{ll} 
  \toprule
       \textbf{Method}  & E$_{tot}$ (Ha)  \\ \hline
       Exact  &  -0.500000 \\ \hline
       TDSE & -0.500000 \\ \hline
       \texttt{Octopus} ($R_{\text{max}} = 50$) & -0.499728 \\ \hline
       \texttt{Octopus} ($R_{\text{max}} = 100$) & -0.501114 \\ 
       \bottomrule
    \end{tabular}
    \caption{Ground-state electronic energy of the hydrogen atom calculated with different methods.}
    \label{tab:E_tot}
\end{table}

\begin{figure*}
\centering
\includegraphics[width=1.0\textwidth]{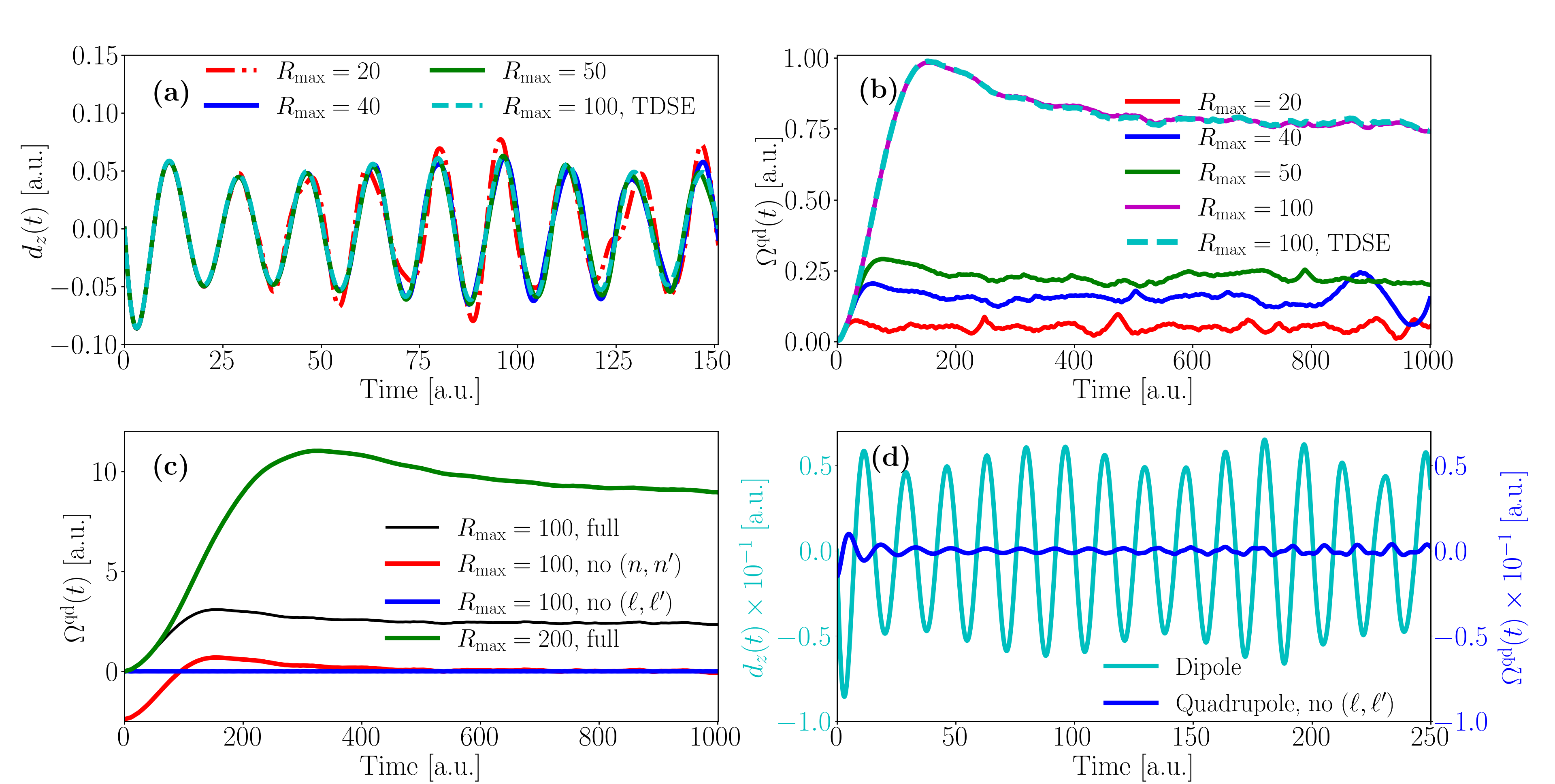}  
\caption{Numerical convergence of the time-dependent observables for a hydrogen atom excited by a spatially uniform, $\delta$-kick electric field with amplitude $E_0 = 0.05$~a.u. across varying simulation box radii $R_{\text{max}}$. 
(a) Induced dipole moment $d_z(t)$ converging at $R_{\text{max}} \ge 40$~a.u. to the reference TDSE result obtained with $R_{\text{max}} = 100$~a.u. 
(b) Time-dependent quadrupole moment $\Omega^{\mathrm{qd}}(t)$, scaling in amplitude with the box size due to the $\sim r^2$ coordinate-dependence of the operator. In this case, convergence to the TDSE reference is achieved at $R_{\text{max}} = 100$~a.u. 
(c) TDSE analysis of the quadrupole moment at $R_{\text{max}} = 100$ and $200$~a.u., comparing the full calculation against channel-deactivated limits where either principal shell couplings [no $(n,n')$] or angular momentum diagonal channels [no $(\ell,\ell')$] are omitted. 
(d) Time-evolution of the converged dipole moment $d_z(t)$ and the purely off-diagonal, angular-momentum-dependent quadrupole moment $\Omega^{\mathrm{qd}}(t)$ [no $(\ell,\ell')$] in the uniform weak-field regime.}
\label{fig.hdp_qd05}
\end{figure*}
We examine the atomic response to a spatially uniform, $z$-polarized $\delta$-kick with amplitude $E_0 = 0.05$~a.u. (WD regime). As shown in Fig.~\ref{fig.hdp_qd05}a, the time-dependent dipole moment $d_z(t)$ converges rapidly, stabilizing for a simulation box radius of $R_{\text{max}} = 50$~a.u. In contrast, the quadrupole moment exhibits a persistent, monotonic increase with the simulation volume (Fig.~\ref{fig.hdp_qd05}b). This behavior is an intrinsic property of the quadrupole operator. For a $z$-polarized field, the active component of the quadrupole operator is $\hat{Q}_{zz} = 3z^2 - r^2$ [Eq.~\eqref{eq.OmeQ}]. According to the selection rules $\Delta\ell = 0, \pm 2$, the expectation value expands as: 
\begin{equation}\label{eq.quad_split}
\Omega^{\rm qd}(t) = \underbrace{\sum_{i} |c_i|^2 \Omega^{\rm qd}_{ii}}_\text{Static} + \underbrace{\sum_{i,j (i \neq j)} c^{*}_i c_j \Omega^{\rm qd}_{ij} e^{i\omega_{ij}t}}_{\text{Dynamic}}.
\end{equation}

 The coefficients $c_i(t)$ are defined in Eq.~\eqref{eq.c_coef_refined} from the expansion of the full time-dependent wavefunction in Eq.~\eqref{eq.psif}. For a complete (untruncated) basis, these (discretized) coefficients satisfy the normalization condition $\sum_j |c_j|^2 = 1$. On a truncated basis, the coefficients are renormalized, which preserves the evaluation of both dipole and quadrupole moments in Eq.~\eqref{eq.quad_split}.
The static term represents the intrinsic quadrupole moment, inducing a rigid vertical shift without altering the dynamic evolution. For the $\Delta\ell = 0$ channels in the dynamic term, the diagonal expectation value $\langle r^2 \rangle$ for a hydrogenic state $|n \rangle$ scales as~\cite{BetheSalpeter1957}):
\begin{equation}
  \langle r^2 \rangle_{n\ell} = \frac{n^2}{2} [ 5 n^2 + 1 - 3 \ell (\ell + 1) ].  
\end{equation}
Since the spatial extent scales asymptotically as $\langle r^2 \rangle \sim n^4$ and the maximum attainable principal quantum number $n$ on a finite grid scales as $n^2 \sim R_{\text{max}}$, the diagonal quadrupole contribution exhibits an explicit quadratic volume scaling:
\begin{equation}\langle r^2 \rangle \sim R_{\text{max}}^2,
\end{equation}
explaining the overall amplitude scaling shown in Fig.~\ref{fig.hdp_qd05}b. Given the quantitative agreement between \texttt{Octopus} and TDSE results, this spatial scaling behavior is investigated within the TDSE framework for physical clarity.

 The static contribution in Eq.~\eqref{eq.quad_split} remains strictly time independent (Fig.~\ref{fig.hdp_qd05}c). The oscillatory components, which encode transition frequencies and spectral features, remain invariant against variations in $R_{\text{max}}$. The $\ell = \ell'$ diagonal channels constitute a rotationally invariant coupling pathway mediating quadrupole transitions between states with different principal quantum numbers ($np \to n'p$) while preserving angular momentum. This mechanism couples high-lying excited states, offering a unique spectral signature that is absent within the dipole approximation.

To isolate the absorption dynamics and compute the cross section, we perform an orbital decomposition analysis that suppresses the static background shifts by explicitly deactivating specific coupling channels (Fig.~\ref{fig.hdp_qd05}c). In the absence of angular-momentum diagonal channels, the oscillatory quadrupole signal centers symmetrically on zero and directly reflects the $\Delta\ell=\pm2$ transition amplitudes. At a uniform field strength of $E_0 = 0.05$~a.u. ($\approx 8.8 \times 10^{13} \text{ W/cm}^2$), this dynamic quadrupole response remains several orders of magnitude weaker than the dipole signal (Fig.~\ref{fig.hdp_qd05}d), confirming the weak-field regime. Nevertheless, these quadrupole oscillations provide a critical diagnostic tool for identifying higher-order coupling channels that become dominant in the SD regime.

\subsection{Dependence of the photoabsorption cross section on the simulation box size}\label{app:cros_tdse_tddft}
We evaluate the impact of the box-size dependence of $\Omega^{\rm qd}(t)$ [Eq.~\eqref{eq.quad_split}] on the absorption cross section. Calculations are performed within the TDSE framework under a spatially non-uniform field gradient $\mathbf{E}(0,0,z)$ with an amplitude of $E_0 = 0.05$~a.u.
Since the field contains only the $z$-component, only $\sigma_{zz}$ couples to the perturbation, rendering orientational averaging unnecessary. 

Since the static diagonal elements act as a rigid shift in the time domain (see Fig.~\ref{fig.hdp_qd05}c), the extracted transition frequencies and spectral shapes remain invariant under Fourier transformation. On the other hand, the dynamical off-diagonal elements in Eq.~\eqref{eq.quad_split} progressively include higher electronic states as the simulation 
\begin{figure}[!t]
\centering
\includegraphics[width=0.5\textwidth]{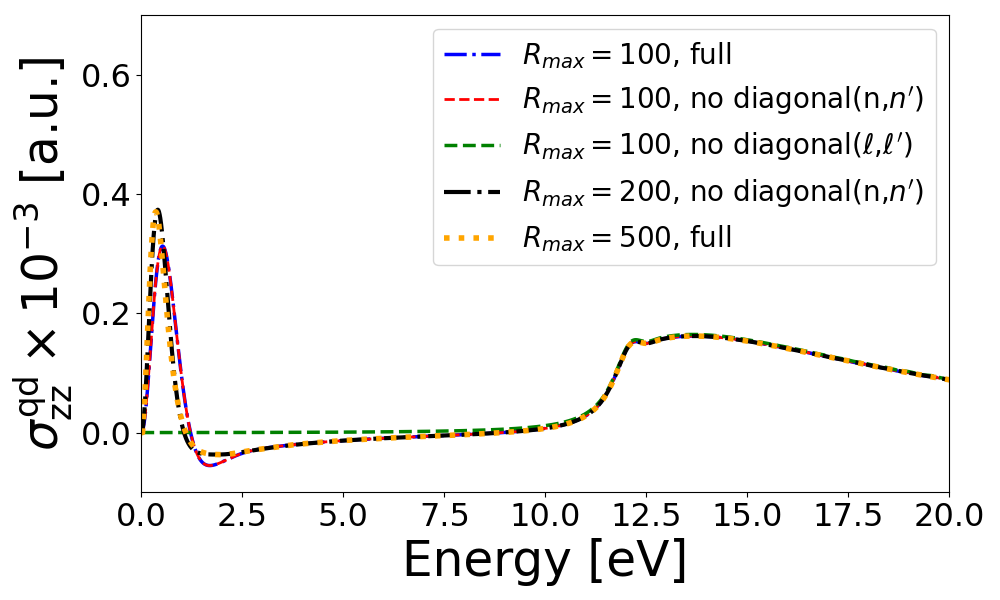}
\caption{Quadrupole cross section ($\sigma^{\mathrm{qd}}_{\rm zz}$) as a function of energy for a field strength $E_0 = 0.05$ a.u., obtained from TDSE calculations using different simulation box radii $R_{\text{max}}$ and combinations of diagonal and off-diagonal elements. }
\label{fig.sig_tdse}
\end{figure}
volume expands. Increasing the box boundary beyond $R_\text{max} = 500$~a.u. leaves the negative spectral features in the cross section completely unchanged, confirming their numerical stability and physical reality. 

 The roles played by different coupling channels appear in both the SD regime and the spatially non-uniform field case. Excluding principal shell diagonal couplings $(n\neq n')$ leaves the cross section essentially unchanged (Fig.~\ref{fig.sig_tdse}). In contrast, excluding angular-momentum diagonal channels $(\ell,\ell')$ isolates pure quadrupole transitions satisfying the selection rule $\Delta\ell = \pm 2$. This confirms that the primary high-energy quadrupole resonance originates exclusively from these angular-momentum-altering pathways.

Conversely, retaining angular-momentum channels $(\ell,\ell')$ restricts transitions to $n \to n'$ excitations within the same angular-momentum subshell. This intra-subshell coupling mechanism governs the sub-eV low-energy response, generating the characteristic positive absorption and negative stimulated emission features seen in Fig.~\ref{fig.sig_tdse}. These low-energy features reflect direct multiphoton and field-assisted transitions connecting closely spaced excited states.

\bibliography{bib.bib}

\end{document}